\documentclass[aps,prd,showpacs,nofootinbib, twocolumn,floatfix,nobalancelastpage]{revtex4}
\usepackage{graphicx}\usepackage{natbib}
\usepackage{times}\usepackage[T1]{fontenc}\usepackage{textcomp}\usepackage{mathptmx}

\begin{document}

\title{Enhanced Cosmological GRB Rates and Implications for Cosmogenic Neutrinos}

\author{Hasan Y{\"u}ksel}
%\email{yuksel@mps.ohio-state.edu}
\affiliation{Department of Physics, Ohio State University, Columbus, Ohio 43210
\\Center for Cosmology and Astro-Particle Physics, Ohio State University, Columbus, Ohio 43210}

\author{Matthew D. Kistler}
%\email{kistler@mps.ohio-state.edu}
\affiliation{Department of Physics, Ohio State University, Columbus, Ohio 43210
\\Center for Cosmology and Astro-Particle Physics, Ohio State University, Columbus, Ohio 43210}

\date{20 February 2007}

\begin{abstract}
Gamma-ray bursts, which are among the most violent events in the universe, are one of the few viable candidates to produce ultrahigh energy cosmic rays.  Recently, observations have revealed that GRBs generally originate from metal-poor, low-luminosity galaxies and do not directly trace cosmic star formation, as might have been assumed from their association with core-collapse supernovae.  Several implications follow from these findings.  The redshift distribution of observed GRBs is expected to peak at higher redshift (compared to cosmic star formation), which is supported by the mean redshift of the Swift GRB sample, $\left<z\right>\sim3$.  If GRBs are, in fact, the source of the observed UHECR, then cosmic-ray production would evolve with redshift in a stronger fashion than has been previously suggested.  This necessarily leads, through the GZK process, to an enhancement in the flux of cosmogenic neutrinos, providing a near-term approach for testing the gamma-ray burst--cosmic ray connection with ongoing and proposed UHE neutrino experiments.
\end{abstract}
% 95.85.Ry     Neutrino, muon, pion, and other elementary particles; cosmic rays
% 98.70.Rz     gamma-ray sources; gamma-ray bursts
% 98.70.Sa     Cosmic rays
\maketitle

%--------------------------------------------------------------------%
\section{Introduction}
The origin of ultrahigh energy cosmic rays (UHECR) is one of the great remaining mysteries in astrophysics~\cite{Gaisser:2005tu}.  The cosmic-ray spectrum has been measured to beyond $10^{19}$~eV, with a number of events with energy exceeding $10^{20}$~eV~\cite{CR-data, Abraham:2004dt}; however, it is still debated how such highly energetic particles can be produced.  It is now generally accepted that UHECR are of an extragalactic origin~\cite{Hillas:1985is}.  However, above $\sim 10^{19.5}$~eV, the process of photopion production ($p \, \gamma \rightarrow N \, \pi$) on the cosmic microwave background (CMB) is expected to lead to a significant diminution of the cosmic-ray spectrum, the well-known GZK effect~\cite{GZK}.  The relatively short attenuation length associated with the GZK process~\cite{Stecker:1968uc, Stanev:2000fb} necessitates that the observed UHECR arise from local sources.

The decay of charged pions produced in this process results in a flux of ultrahigh energy \textit{cosmogenic} neutrinos~\cite{GZKnu, Hill}.  While the observed UHECR spectrum is somewhat insensitive to variations in cosmic source evolution~\cite{DeMarco:2005ia}, the cosmogenic neutrino flux can be greatly enhanced by strong evolution with redshift~\cite{Hill, ESS}, as neutrinos can be produced in larger quantities due to the decreased photopion threshold (since $T_{\rm CMB} \propto 1+z$), and can themselves traverse cosmological distances without attenuation.

Few classes of astrophysical objects can possibly account for the observed cosmic-ray spectrum~\cite{Hillas:1985is}.  Active galactic nuclei (AGN) have long been considered as possible UHECR sources~\cite{AGN}. Relatively recently, a potential connection between gamma-ray bursts (GRBs) and UHECR has been explored~\cite{WaxmanGRB-CR, Vietri:1995hs, Dermer:2000yd}.  GRBs, which are now generally accepted to be related to core-collapse supernovae~\cite{Stanek:2003tw}, are violent events which release great amounts of energy in the form of gamma-rays ($\sim 10^{51}-10^{52}$ erg)~\cite{Paczynski:1986px}.  A number of models have been proposed to utilize their ultra-relativistic environment to accelerate protons to energies $\gtrsim 10^{20}$~eV~\cite{WaxmanGRB-CR, Vietri:1995hs, Dermer:2000yd}.  It has also been noted that the source emissivity required to account for $\gtrsim 10^{19}$~eV cosmic rays is comparable to that of gamma-ray bursts~\cite{Waxman:1995dg, WaxmanGRB-CR}.
%
%%%%%%%%%%%%%%%%%%%%%%%%%
\begin{figure}[b]
\includegraphics[width=3.25in,clip=true]{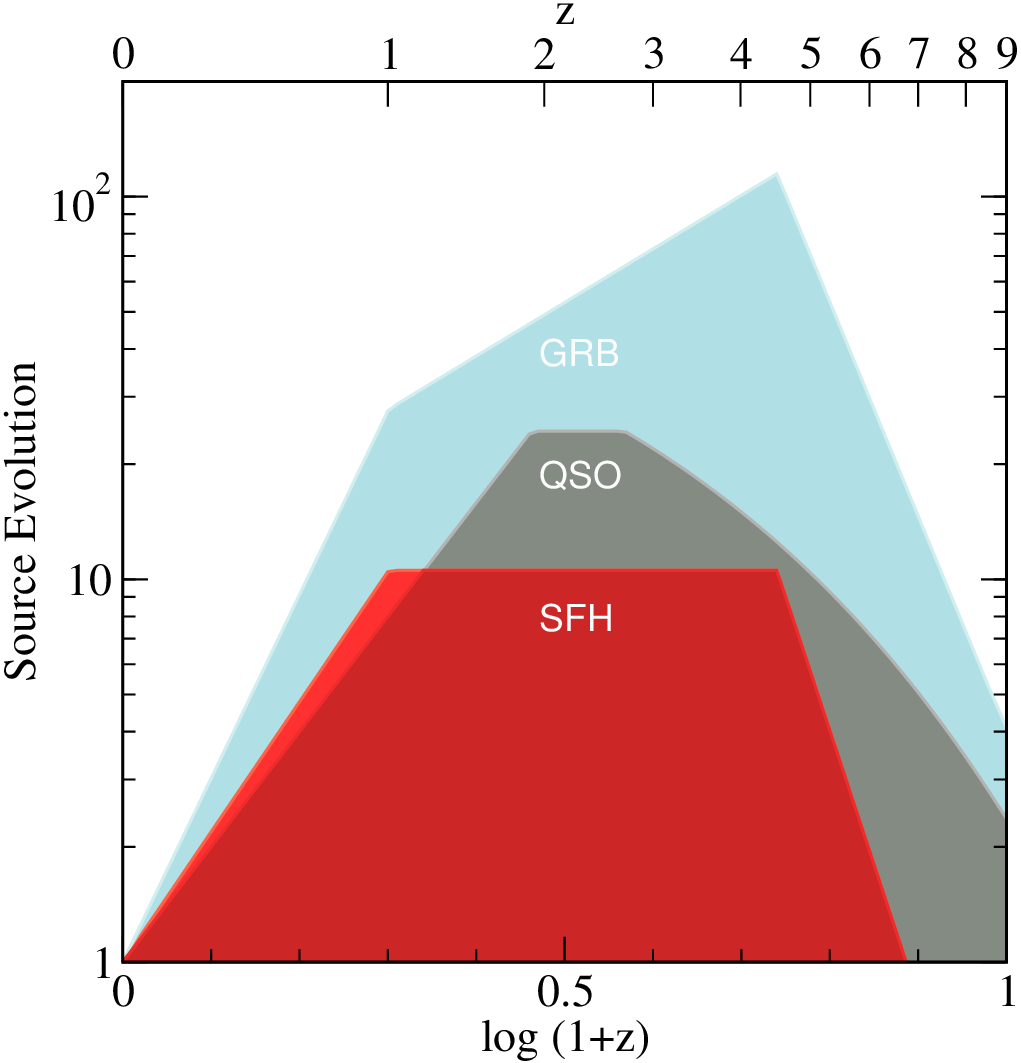}
\caption{Models of cosmic-ray source evolution (i.e.~yield vs.~$z$, normalized to 1 at $z=0$).  From top-to-bottom, the metallicity-dependent GRB rate density (this work), the quasar (QSO) evolution model used in Refs.~\cite{Waxman:1998yy, ESS}, and the fit to the cosmic star formation history (SFH) of Ref.~\cite{Hopkins:2006bw}.  Models similar to the latter two have been frequently used in cosmic-ray studies.
\label{fig:cr_evol}}
\end{figure}
%%%%%%%%%%%%%%%%%%%%%%%%%

Just as the core-collapse supernova rate density seems to follow the cosmic star formation history (SFH)~\cite{Hopkins:2006bw}, the same might be expected of the cosmological GRB rate density~\cite{Totani:1998zs}.  However, there is mounting evidence that GRBs are \textit{not} an unbiased tracer of the SFH~\cite{LeFloc'h:2003yp, LeFloc'h}.  In particular, the host galaxies of GRBs have a distinct tendency to be subluminous~\cite{LeFloc'h:2003yp} and metal-poor~\cite{HOST-METAL}.  This has been demonstrated for GRB hosts both locally~\cite{Stanek:2006gc} and at cosmological distances~\cite{Fruchter}, which suggests that low metallicity is a key ingredient in the production of a gamma-ray burst~\cite{Stanek:2006gc}.  As we discuss in Section~\ref{RateSect}, a rapidly rotating star, as required in the collapsar model~\cite{Woosley:1993wj}, can retain much of its original mass and angular momentum if it is metal-poor~\cite{Yoon:2005tv}.  An anti-correlation with metallicity would imply that the cosmological GRB rate peaks at a higher redshift~\cite{Langer:2005hu, Yoon:2006fr}, which now seems to be indicated by Swift observations~\cite{Le:2006pt, Daigne:2006kf}.  Simply put: the metallicity of the universe \textit{decreases} with redshift, which implies a \textit{stronger} evolution of the GRB rate density than would be expected from the SFH alone.

When GRBs are considered as the source of UHECR (with identical cosmic-ray production per burst), the change in the cosmological cosmic-ray emissivity is simply determined by the burst rate history. In the metallicity-biased GRB model, this evolution is quite strong, as illustrated in Fig.~\ref{fig:cr_evol}, even exceeding that of models used in past cosmic-ray studies, which have traced quasar (QSO) luminosities~\cite{QSO} or the SFH.  We examine the effect of enhanced GRB rate evolution on the expected flux of cosmogenic neutrinos, the measurement of which may provide the only way to break degeneracies between cosmic-ray models~\cite{Seckel:2005cm}.   In addition to naturally explaining the abundance of high-redshift bursts observed by Swift, this strong evolution leads to a measurable neutrino signal, improving the near-term prospects for assessing this scenario with upcoming ultrahigh energy neutrino detectors~\cite{Barwick:2005hn, ASK, AugerNU}. This result would still hold, in general, even if another mechanism is ultimately shown to account for the Swift results through enhanced GRB evolution.  Though we focus on the impact on cosmogenic neutrino production (as Dermer and Holmes have also recently done for GRB rates directly tracing several SFHs~\cite{Dermer-GZK}), the metallicity-biased evolution model would also have implications for predictions of diffuse neutrino fluxes directly produced in GRBs (e.g., Refs.~\cite{Dermer:2000yd, GRB-NU}) and their prospects for detection~\cite{Ahrens:2003ix, Cuoco:2006qd}.

%--------------------------------------------------------------------%
\section{Metals and the Predicted GRB Rate}
\label{RateSect}
Any treatment of cosmological cosmic-ray or neutrino production must account for source evolution.  Traditionally, the most commonly used cosmic-ray evolution models have tracked quasars~\cite{QSO} or the SFH~\cite{Hopkins:2006bw} (particularly in GRB-related studies).  Observations indicate, however, that GRBs do not faithfully trace the SFH~\cite{LeFloc'h:2003yp, LeFloc'h}.  In fact, it now appears that the GRB rate density is actually evolving more strongly than the standard SFH~\cite{Le:2006pt, Daigne:2006kf, GRB-LUM}.  A natural explanation may be found by considering the properties of GRB host galaxies in the context of the single-star collapsar model~\cite{Woosley:1993wj, Yoon:2005tv}.  In this model, GRB progenitors are rapidly-rotating Wolf-Rayet stars which undergo a core-collapse event that produces a black hole (or possibly a rapidly-rotating neutron star)~\cite{Heger:2002by}.  After collapse, this rapid rotation allows for the formation of highly relativistic jets which, when viewed on-axis, are observed as a burst~\cite{Zhang:2002yk}.

Observationally, all known supernova counterparts of GRBs are Type Ic, with the implication that the dying star lacked an outer hydrogen/helium envelope~\cite{Stanek:2003tw}.  The winds of Wolf-Rayet stars, which are typically the cause of the loss of this envelope, are known to increase in strength with stellar metallicity (particularly iron)~\cite{WINDS}.  Importantly, in this wind-induced loss process, angular momentum (which is particularly important for forming jets~\cite{Rockefeller:2006ep}) is lost along with mass~\cite{Petrovic:2005bq}.  This loss of angular momentum can be avoided if the progenitor has a very low metal content.  In addition to having weaker winds, a rapidly-rotating, metal-poor massive star can avoid the production of an envelope altogether by completely mixing its interior, which results in the hydrogen being circulated into the core and burned~\cite{Yoon:2005tv}.  This would be impossible if the star was not metal-poor, as stellar mixing is expected to be inhibited by a high metal content~\cite{Heger:1999ax}.

Studies of GRB hosts demonstrate that these galaxies tend to have a low metallicity~\cite{Stanek:2006gc, Fruchter}.  Compared to considerations of galactic luminosity observations alone~\cite{Wolf:2006uz}, direct spectroscopic measurements reveal metallicities that are significantly lower than expected~\cite{Stanek:2006gc}.  While the metallicity of the GRB progenitor may not be directly measured, the metallicity of the galaxy itself should be indicative.  A connection between GRBs and metallicity implies that the cosmological GRB rate should be dependent upon the formation rate of metal-poor stars.  At higher redshifts, the universe is less metal-enriched than at the present~\cite{Jimenez:2006ea}, resulting in an increase in the expected evolution of the GRB rate density (compared to the SFH), as shown by Langer and Norman~\cite{Langer:2005hu} (see also Refs.~\cite{Jakobsson:2005jc, Natarajan:2005tp}).  Accounting for stellar evolution effects, Yoon, Langer, and Norman calculated the expected ratio of GRBs to core-collapse supernovae as a function of redshift, as seen in Fig.~7 of Ref.~\cite{Yoon:2006fr}, which we approximate as $\dot{n}_{\rm GRB}(z) \propto (1+z)^{1.4} \dot{n}_{\rm SN}(z)$.  This is consistent with estimates of this ratio~\cite{Podsiadlowski:2004mt} and can be used to calculate the increase in the absolute GRB rate density (of importance to cosmic-ray studies).

As the progenitors of core-collapse supernovae are very short lived, $\dot{n}_{\rm SN}(z)$ is expected to closely follow the SFH (assuming an unchanging IMF~\cite{Salpeter:1955it}).  Utilizing constraints from the diffuse supernova neutrino background~\cite{dsnb-nu} (in addition to direct observations), an updated SFH was derived by Hopkins and Beacom~\cite{Hopkins:2006bw}, which is well-described by the piecewise-linear fit: $\dot{n}_{\rm SFH}(z) \propto (1+z)^\alpha$, where $\alpha =$ (3.4, 0, -7) when ($z<1$, $1<z<4.5$, $4.5<z$).  Combining this SFH with the parametrized form of the GRB/SN ratio, we find the source evolution term, $\mathcal{W}_{\rm GRB}(z)$, to have the form
\begin{equation}
\mathcal{W}_{\rm GRB}(z) \propto
\begin{array}{lcl}
(1 + z )^{4.8}  		& {\rm :}   	&  z  <  1 \\
(1 + z)^{1.4}			& {\rm :}   	&  1< z < 4.5\\
(1 + z)^{-5.6}			& {\rm :}   	&  4.5< z \,,
\end{array}
\label{eq} 
\end {equation}
with $\dot{n}_{\rm GRB}(z) = \dot{n}_{\rm GRB}(0) \times \mathcal{W}_{\rm GRB}(z)$.
In Fig.~\ref{fig:cr_evol}, we present this evolution model, in comparison to the SFH alone, as well as the quasar (QSO) evolution model used in Refs.~\cite{Waxman:1998yy, ESS}.  Note that our rate history evolves as $\mathcal{W}_{\rm GRB}(z)\sim (1+z)^{4.8}$ for $z<1$, which is substantially steeper than the models commonly considered.  In particular, the QSO model, which has been extensively used, only rises as $\mathcal{W}_{\rm QSO}(z)\sim (1+z)^3$.  Previous studies have found an evolution in GRB luminosity $\propto (1+z)^{1.5}$~\cite{GRB-LUM}, which would result in a similar effect on cosmic-ray evolution.  While this may just be due to stronger beaming, we will not consider this scenario and instead focus only  upon changes in the GRB rate density.  GRB rates that evolve more strongly than the standard SFH have, in fact, recently been determined using Swift data~\cite{Le:2006pt, Daigne:2006kf}.

To directly check for compatibility with observations, we calculate the expected redshift distribution (assuming Swift~\cite{Burrows:2005gf} detection sensitivity) for the metallicity-dependent evolution, following the GRB model presented by Le and Dermer~\cite{Le:2006pt}, and compare to a sample of 46 long bursts discovered by Swift (compiled from the updated lists of Ref.~\cite{Jakobsson:2005jc}), all of which have reliable redshifts. Details of this calculation are contained in Appendix~\ref{appA}.  In particular, we assume an opening angle that varies between $\theta_{\rm min}=0.05$ and $\theta_{\rm max}= 0.5$ (with a power law distribution of the form $\propto (1-\cos \theta)^s$ where the index is $s\sim-1.5$) and an absolute gamma-ray energy per GRB of $\epsilon_{\gamma} = 5 \times 10^{51} \rm erg$ released in $\sim \delta t = 10$~s, all of which are well within the reasonable ranges of these parameters~\cite{Ghirlanda:2006ax} assuming a uniform jet (see Ref.~\cite{Rossi:2001pk} for alternative jet models). 

\begin{figure}[t]
\includegraphics[width=3.25in,clip=true]{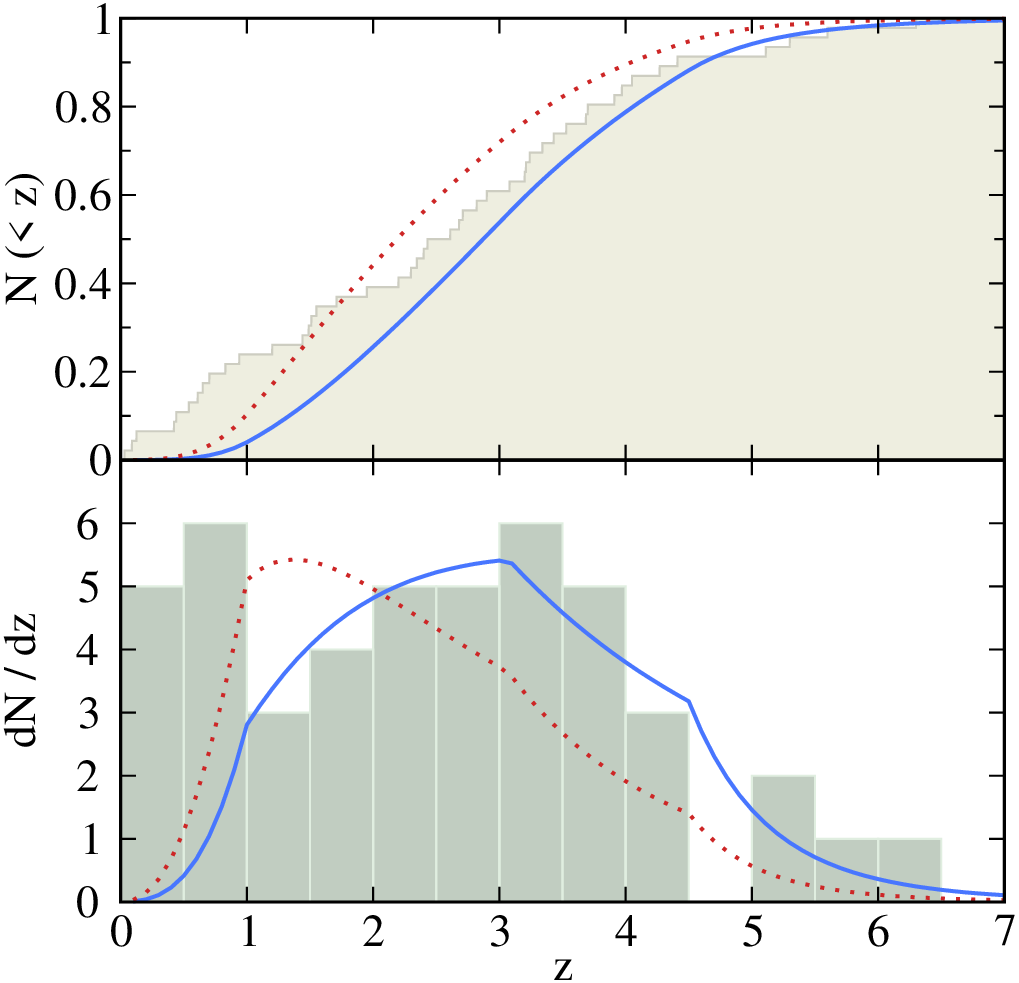}
\caption{Predicted redshift distributions of observable gamma-ray bursts from the strongly-evolving (metallicity-biased) model (solid line) and the SFH alone (dotted line), compared to Swift bursts with known redshifts.  In the top panel, the cumulative burst distribution is shown (shaded region).  The bottom panel shows the differential distributions (with arbitrary normalizations), which allow for a direct comparison with the data (in bins of width $0.5$).  The excess of low-$z$ events may be due to a LLGRB class (as discussed in Section~\ref{RateSect}).
\label{fig:cr_sw}}
\end{figure}

In Fig.~\ref{fig:cr_sw}, we compare the distribution of Swift bursts to the predictions of the metallicity-biased model and the standard SFH.  The top panel suggests that the cumulative distribution of observed bursts (shaded region) can be approximately described by either the metallicity-dependent evolution or the SFH.  However, another perspective is presented in the bottom panel, where the differential event distribution is shown (in redshift bins of width 0.5).  This allows for a more direct comparison with redshift data, which reveals that most bursts are discovered between $z \sim 1-4$, in good agreement with our strongly evolving rate.  This stronger evolution allows for a better fit to the data, when compared to the SFH.  A more detailed discussion of the uncertainties involved in GRB detection (particularly the observed bursts that lack a reliable redshift), as well as the various degeneracies that exist between parameters in GRB modeling, is beyond the scope of this work.  We refer the interested reader to Ref.~\cite{Le:2006pt} (and references therein).

%%%%%%%%%%%%%%%%%%%%%%%%%
%%%%%%%%%%%%%%%%%%%%%%%%%

While the metallicity-biased model is in generally good agreement with Swift observations at high redshifts, an excess of events seems to be present at low redshifts, which affects the cumulative distribution.  This might be explained by a separate class of low-luminosity GRBs (LLGRB)~\cite{Norris:2002rq, Liang:2006ci}.  While these events are typically weaker than cosmological bursts, they may be several orders of magnitude more abundant in the local universe~\cite{GRB060218}.  One possibility is that these LLGRBs are the result of progenitors that had a higher metal content than typical GRBs, leading to a less luminous (and less beamed) gamma-ray output.  In the context of the collapsar model, this is not an unreasonable conclusion, as higher progenitor metallicity should lead to increased loss rates of both mass and angular momentum.

One consequence of the strong evolution that we present is a predicted burst rate density of $\dot{n}_{\rm GRB}(z=0) \sim 4$ Gpc$^{-3}$ yr$^{-1}$ in the local universe (when normalized to the observed redshift distribution), which quickly rises to $\sim 20$ Gpc$^{-3}$ yr$^{-1}$ by redshift $z \sim 0.4$.  The average emissivity from these bursts (at $z \lesssim 0.4$) is  $\mathcal{E}_{\rm GRB} \sim \epsilon_{\gamma} \times 10$ Gpc$^{-3}$ yr$^{-1} \sim 5 \times 10^{43}$ erg Mpc$^{-3}$ yr$^{-1}$, which is comparable to the emissivity required to account for the $\gtrsim 10^{19}$~eV UHECR flux, $\mathcal{E}_{\rm CR} \sim \mathrm{few} \times 10^{44}$ erg Mpc$^{-3}$ yr$^{-1}$ (as found in Ref.~\cite{WaxmanGRB-CR}).  Considering that there is no \textit{a priori} reason for these numbers to be so similar, along with the fact that the region of the cosmic-ray spectrum of greatest interest ($\gtrsim 10^{19}$~eV) must arise from sources at $z \lesssim 0.4$ (as we will discuss in Section~\ref{UHECR}), this result is quite interesting.  GRBs may also be more luminous in cosmic rays due to a baryon loading factor ($f_{\rm CR}$) that may be $\gtrsim 10$~\cite{Dermer:2006bb, Dermer:1998ad}, which could account for any difference.  These considerations, along with the isotropy of the measured cosmic-ray spectrum, allow for GRBs to be further examined as a candidate to produce the observed UHECR.

%--------------------------------------------------------------------%
\section{UHECR and GRBs}
\label{UHECR}
To be considered as a viable source of UHECR, gamma-ray bursts must have the ability to both accelerate protons to energies $\gtrsim 10^{20}$~eV and generate a cosmic-ray flux adequate to explain the observed spectrum.  In conventional GRB models, a portion of the kinetic energy of a relativistically expanding fireball (with Lorentz factor $\Gamma \sim$ few hundred) is converted into internal energy~\cite{Piran:1999kx}.  Electrons are accelerated inside this jet by internal shocks and subsequently produce gamma-rays through synchrotron and inverse-Compton processes~\cite{INTERNAL}.

Protons should also be shock-accelerated in a similar fashion.  In the internal shock model, the shocks that accelerate protons are expected to be only mildly relativistic in the wind rest frame, resulting in an $\sim E^{-2}$ spectrum.  In order to efficiently accelerate protons to ultrahigh energies, the time scale of acceleration should be shorter than both the wind expansion time (to allow for an adequate period of confinement in shocked regions) and the proton energy loss time scale.  The former sets the ratio of magnetic field and electron energy densities to order unity,  which is necessary in order to account for gamma-ray emission from synchrotron emission boosted to the observer's frame.  The latter imposes an upper limit on magnetic field strength (and lower limit on Lorentz factor).  See Refs.~\cite{WaxmanGRB-CR, Vietri:1995hs, Dermer:2000yd} for details (and alternative models).

Cosmic rays mainly lose energy through three processes during propagation.  At very high energies ($ \gtrsim 3\times$10$^{19}$~eV), UHECR energy loss is dominated by photopion production on the CMB, $p \, \gamma \rightarrow N \, \pi$, which has a maximum cross section at the $\Delta$(1232) resonance~\cite{Stecker:1968uc, Mucke:1998mk}.  Below the photopion threshold (and with $E_p \gtrsim 10^{18}$~eV), electron-positron pair production on diffuse photons, $p \, \gamma \rightarrow p\, e^+e^-$, becomes important~\cite{Blumenthal:1970nn}.  While the cross section for this process is high, each interaction results in only a small energy loss.  Finally, the expansion of the universe results in adiabatic energy loss, which is independent of energy. 

We can account for the total proton energy loss through the characteristic timescales associated with each process (as in Refs.~\cite{Waxman:1995dg,Berezinsky:2002nc}): $\tau_T^{-1}(E_p,z) = \tau_\pi^{-1}(E_p,z) + \tau_{\rm pair}^{-1}(E_p,z) + \tau_a^{-1}(z)$.  The rate of energy loss for propagating protons is then $ d \ln{E_p}/dt = \tau_T^{-1}(E_p,z)$.  With the redshift-dependent energy loss rate determined, the injection energies of cosmic rays, $E^\prime_p=E^\prime_p(E_p,z)$, can be calculated as a function of detected energy ($E_p$) and originating redshift ($z$) through the differential equation
\begin{equation}
\frac{1}{E_p}\frac{dE_p}{dz}=\frac{1}{\tau_T(E_p,z)}\frac{1}{dz/dt}
\label{eq:eloss}
\end{equation}
where ${dz}/{dt} = H_0 (1+z) [\Omega_M (1+z)^3 +\Omega_\Lambda ]^{1/2}$ (with $\Omega_{M} = 0.3$, $\Omega_{\Lambda} = 0.7$, and ${H}_{0} = 70$ (km/s)/Mpc). 

\begin{figure}[t]
\includegraphics[width=3.25in,clip=true]{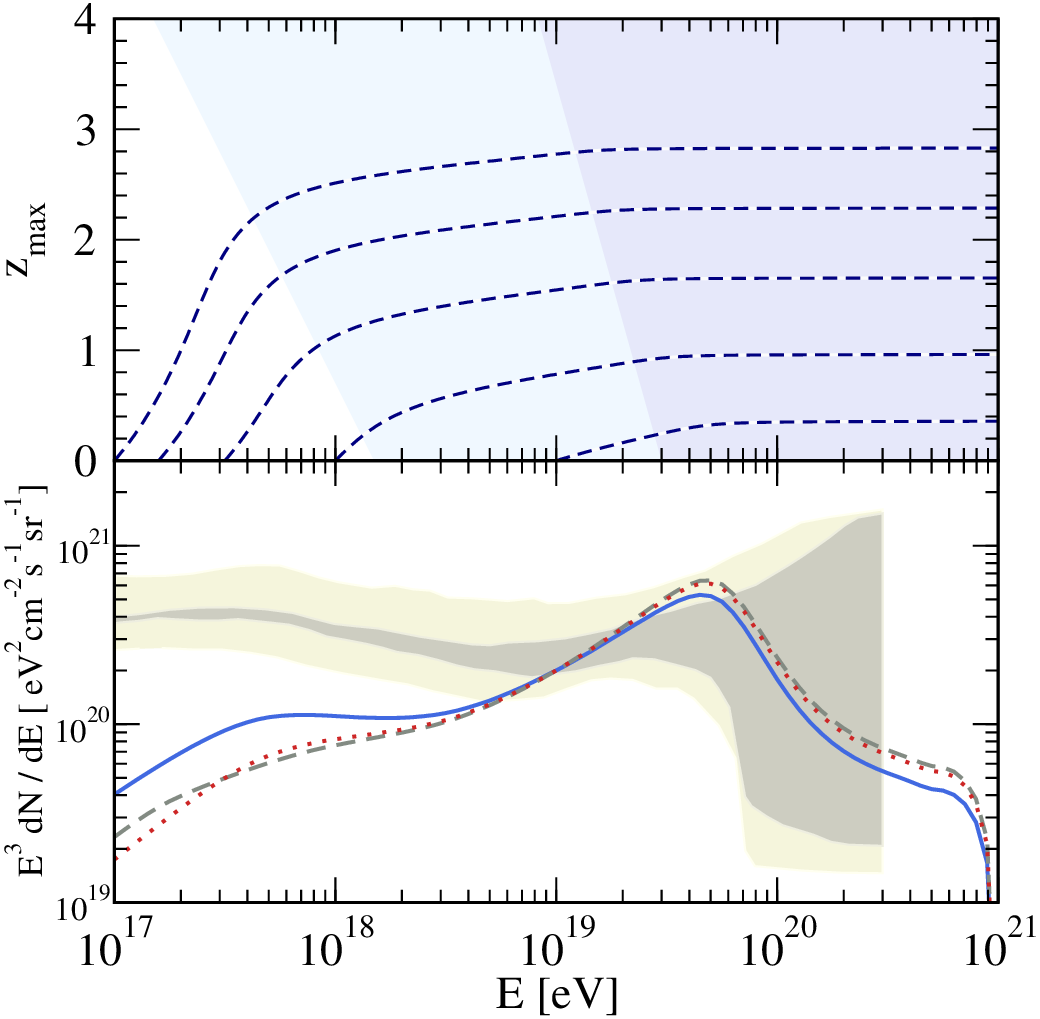}
\caption{Top: Cosmic-ray energy loss with redshift.  Lines illustrate the injection energy required at redshift $z$ in order to be detected at a given energy at $z=0$.  Shown for illustrative purposes are the realms of photopion (dark-shaded region) and pair-production (light-shaded region) losses.
Bottom: Cosmic-ray spectra expected from the GRB (solid), QSO (dashed), and SFH (dotted) source evolution models, assuming $\gamma=2$ and normalization at $E=10^{19}$~eV.  All curves are well within range of experimental data above $10^{19}$~eV (light-shaded region, dark-shaded when normalized to the spectral dip).
\label{fig:cr_eloss}}
\end{figure}

The top panel of Fig.~\ref{fig:cr_eloss} shows cosmic-ray energy loss with redshift.  The lines correspond to the injection energy needed at redshift $z$ in order for a cosmic ray to be observed with a given energy at Earth ($z=0$).  For example, a cosmic-ray proton with a measured energy of $10^{19}$~eV must have been produced at $z < 0.4$.  This effect is analogous to the Fazio-Stecker relation for gamma rays~\cite{FS}.  As energy losses above 3$\times$10$^{19}$~eV are quite severe, the observed UHECR must be produced in the local universe.  This, along with the necessary conditions required to produce UHECR at all, strongly constrains the population of prospective sources.

%%%%%%%%%%%%%%%%%%%%%%%%%
%%%%%%%%%%%%%%%%%%%%%%%%%

In evaluation of the cosmic-ray spectrum,  we have assumed that, at these energies, the cosmic-ray spectrum is entirely composed of protons with an injection spectrum $\varphi(E_p^\prime) =  {\cal N}  E_p^{\prime-\gamma}$ (per unit comoving volume per unit energy, per unit time), which is cut-off at a chosen $E_{\rm cut}$ and normalized to account for the observed spectrum of UHECR. Cosmic rays injected at $E_p^\prime$, will experience energy losses and be detected at $E_p$.  Taking into account the evolution of sources, $\mathcal{W}(z)$, we can calculate the UHECR spectrum as,
\begin{equation}
\frac{dN_p}{dE_p}  =  \frac{c}{4 \pi} \int_0^{z_{max}} \varphi(E_p^\prime) 
\frac{\partial E_p^\prime}{\partial E_p} \frac{\mathcal{W}(z)} {dz/dt} dz \,.
\end{equation}
We calculate the partial derivative ${\partial E_p^\prime(E,z)}/{\partial E_p}$ numerically from Eq.~\ref{eq:eloss}.  In the bottom panel of Fig.~\ref{fig:cr_eloss}, we present the expected cosmic-ray spectrum for the three source evolution models presented in Fig.~\ref{fig:cr_evol}, with an injection spectrum of the form $E^{-2}$ (plotted as  $E^3 \times Flux$ to emphasize spectral features~\cite{Berezinsky:1988wi}).  We choose the normalization of $\varphi$, ${\cal N}$, such that $E^3 \times dN_p/dE_p= 2\times 10^{20}$ eV$^2$ cm$^{-2}$ s$^{-1}$ sr$^{-1}$ at 10$^{19}$~eV for these spectra (and in our subsequent results), as the uncertainties at higher energies await the resolution that will be delivered by Auger~\cite{Abraham:2004dt}. This usually necessitates the local emissivity of sources between $10^{19}-10^{21}$~eV to be on the order of $\mathcal{E}_{\rm CR} \sim 5 \times$10$^{44}$ erg Mpc$^{-3}$ yr$^{-1}$~\cite{Waxman:1995dg, DeMarco:2005ia}.

Note that the data from the AGASA, HiRes and Yakutsk experiments~\cite{CR-data} differs by almost a factor of 2 in overall normalization in our $E^3$ plot (light shaded region), however, the overall spectra do not significantly disagree~\cite{DeMarco:2003ig} and model-independent information can be extracted from the shape alone, especially when various experimental data are normalized to the spectral dip feature at $\sim 10^{19}$~eV~\cite{Berezinsky:2002nc} (dark shaded region).

When considering the cosmic-ray spectrum resulting from GRB evolution, we have followed the general form of Waxman by assuming that the entire UHECR spectrum above $10^{19}$~eV can be accounted for by GRBs, with an $E^{-2}$ injection spectrum~\cite{Waxman:1995dg, WaxmanGRB-CR}.  However, for an injection spectrum of this form, it is challenging to account for cosmic rays with energies $\lesssim 10^{19}$~eV~\cite{Scully:2000dr}.  Strong source evolution with redshift offers a certain degree of assistance.  It has been assumed previously that an extension of the Galactic cosmic-ray spectrum may account for any remaining deficiency~\cite{Bahcall:2002wi}.  Contributions from extragalactic sources with a lower energy reach may also be considered (e.g., AGN, cluster shocks~\cite{CLUSTERS}, or LLGRBs).  If LLGRBs are able to produce cosmic rays (as suggested in Ref.~\cite{LLnu}), they may be able to contribute to the flux of lower energy cosmic rays (since they have a high local rate~\cite{GRB060218}) and alleviate the need for an extension of the Galactic component.  A recent Galactic GRB has also been proposed as a potential source of lower energy cosmic rays~\cite{MW-GRB}.  While the Milky Way is generally metal-rich (contrary to observations of typical GRB hosts), it possesses a gradient in its metallicity distribution which may have allowed for a low-luminosity burst in the past.  Alternatively, one can consider an $E^{-2.2}$ injection spectrum (as might be expected from relativistic shocks), which may offer a better fit at lower energies~\cite{MW-GRB, Dermer-GZK}.  This would require a larger $f_{CR}$, which itself may explain GRBs with low radiative efficiencies~\cite{MW-GRB}.

It is important to note that the cosmic-ray spectrum itself depends only mildly upon evolution (since the observed UHECR with $E\gtrsim10^{19}$~eV must be produced at $z \lesssim 0.4$) and may be explained by a combination of GRBs and a lower-energy component.  However, the strong evolution implied by the metallicity-biased GRB model results in an increased flux of cosmogenic neutrinos.  This neutrino flux, which must be present if GRBs are to account for the observed UHECR, provides an independent test of the GRB--UHECR model, which we will now examine.

%--------------------------------------------------------------------%
\section{Implications For Cosmogenic Neutrinos}
\label{NU}
The flux of cosmogenic neutrinos produced via the GZK process is quite sensitive to cosmic-ray source evolution~\cite{Seckel:2005cm}.  This is due to the unique ability of neutrinos to propagate through large distances at very high energies without appreciable energy loss (unlike cosmic rays), combined with a lower threshold for photopion production at larger $z$ (since $T_{\rm CMB} \propto 1+z$).  In fact, a significant portion of the expected neutrino flux originates between from $z\sim 1-4$.
  
Near the photopion production threshold, about 20\% of the original proton energy is lost in each interaction.  Neutrinos are subsequently produced through the decay chain, $\pi^+ \rightarrow \mu^+ \, \nu_\mu \rightarrow e^+ \, \nu_e \, \overline\nu_\mu \, \nu_\mu$, with each daughter neutrino ($\nu_e,\nu_\mu,\bar\nu_\mu$) receiving $\sim$1/20 of the parent proton energy~\cite{Berezinsky:2005ng}.  At higher energies, multi-pion production dominates~\cite{Mucke:1998mk}; however, this approximation remains viable, as the inelasticity of each interaction increases (approaching 50\%).
%
%%%%%%%%%%%%%%%%%%%%%%%%%%%
\begin{figure}[b]
\includegraphics[width=3.25in,clip=true]{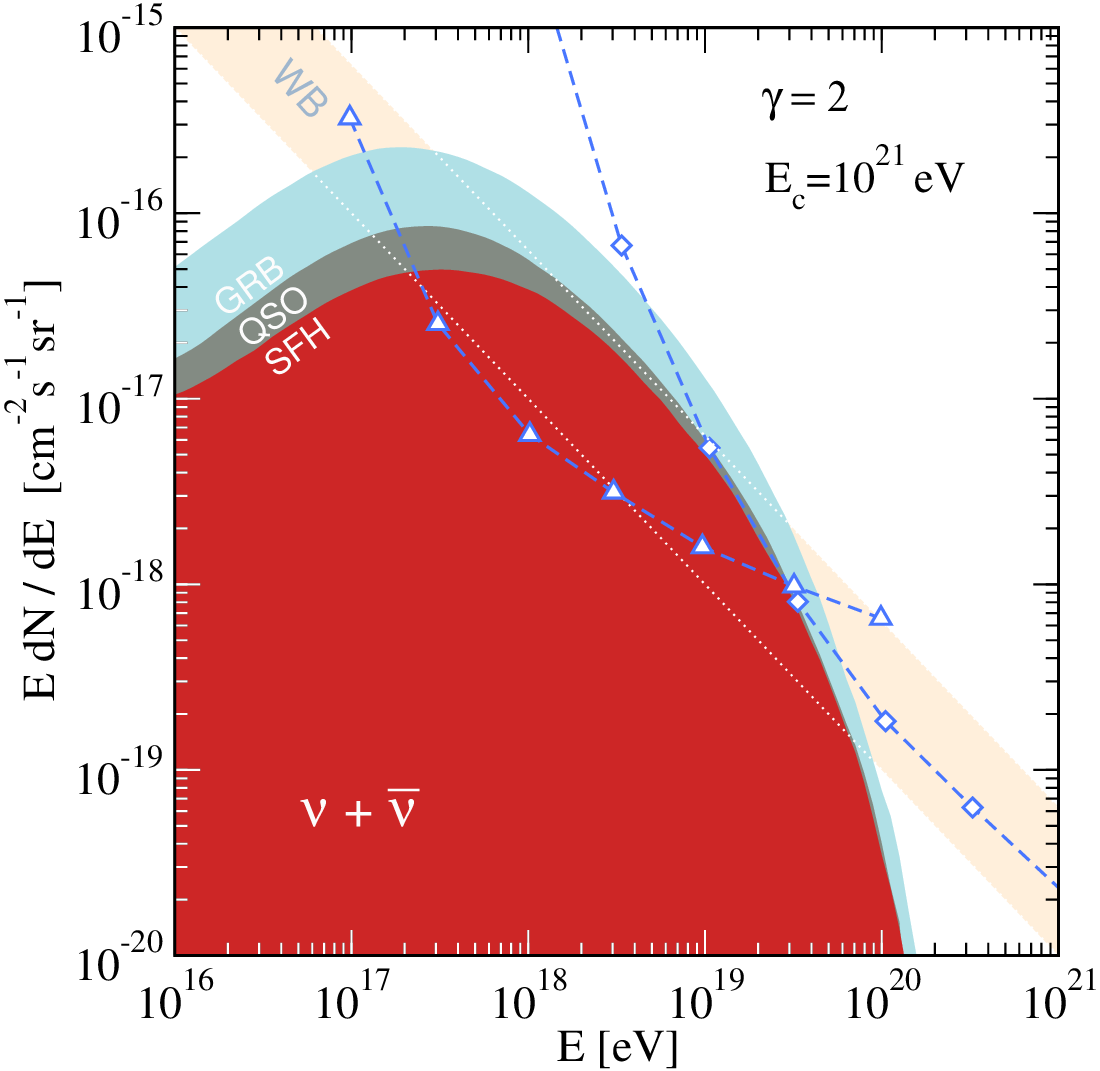}
\caption{Expected (all-flavor) cosmogenic neutrino fluxes assuming various evolution scenarios.  From top-to-bottom, are the fluxes resulting from the strongly evolving (metallicity-dependent) GRB rate density, QSO-like evolution, and the SFH.  Shown for comparison are the Waxman-Bahcall bound (shaded band) and the expected sensitivities for ANITA (diamonds) and ARIANNA/SalSA (triangles).
\label{fig:gzk1}}
\end{figure}
%%%%%%%%%%%%%%%%%%%%%%%%%%%

\begin{figure}[t]
\includegraphics[width=3.25in,clip=true]{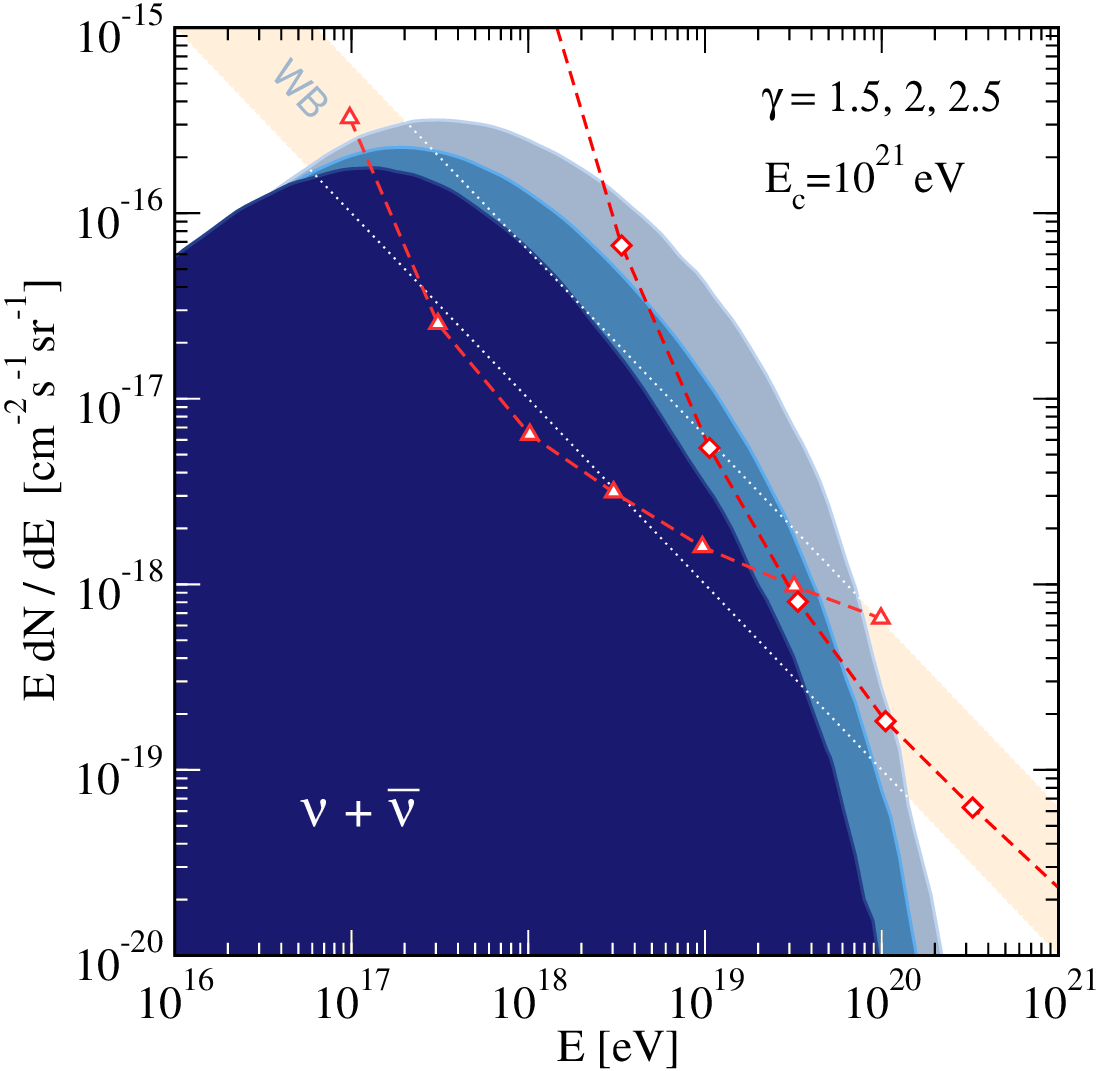}
\caption{Expected (all-flavor) cosmogenic neutrino fluxes resulting from various UHECR injection spectra and assuming 
strong (metallicity-dependent) GRB evolution.  From top-to-bottom, are the fluxes with spectral indices 1.5, 2, and 2.5.  Shown for comparison are the Waxman-Bahcall bound (shaded band) and the expected sensitivities for ANITA (diamonds) and ARIANNA/SalSA (triangles).
\label{fig:gzk2}}
\end{figure}

Waxman and Bahcall (WB) have presented an upper bound on cosmogenic neutrino production (shown in Figs.~\ref{fig:gzk1} and~\ref{fig:gzk2} as a shaded band) based on the assumption of an $E^{-2}$ injection spectrum, with normalization chosen between $10^{19}$~eV to $10^{21}$~eV to produce the observed cosmic-ray spectrum~\cite{Waxman:1998yy}.  This yields an energy-dependent rate of cosmic-ray generation  of ${\cal N}_{\rm WB} \sim 10^{44}$ erg Mpc$^{-3}$ yr$^{-1}$ (with $\mathcal{E}_{\rm WB} \sim 5\times {\cal N}_{\rm WB}$).
The total $\nu_\mu+\bar\nu_\mu $ energy flux at Earth (not corrected for oscillations) is estimated as 
\begin{equation}
E_\nu^2 \frac{dN_{\nu}}{dE_\nu} \approx   \frac{c}{4 \pi} {\cal N}_{\rm WB} \frac{1}{4} \, \xi \, t_H \, 
\approx \xi \times \, 15 \frac{\mathrm{eV} }{\mathrm{cm}^{2} \, \mathrm{s} \, \mathrm{sr}}, 
\end{equation}
where $t_H\approx 10^{10}$ yr is the Hubble time and the factor $1/4$ arises from the assumption that only one quarter of the energy lost is carried away by muon neutrinos. Adiabatic redshift losses and the effect of source evolution are taken into account by $\xi$ (estimated to be $\xi \sim$ 0.6 with no source evolution, $\sim$ 3 with QSO-like evolution)~\cite{Waxman:1998yy}.

The neutrino spectrum produced in the GZK process may be better approximated for a general cosmic-ray injection spectrum and source evolution through a somewhat more sophisticated approach.  As the energy loss distance at these energies is relatively short, we assume that cosmic rays lose all of their energy rapidly.  The fraction of the original proton energy that is lost to neutrinos can then be parametrized with a gradual step function, ${\cal S}\!(E)={0.45}/(1+(E_t/E)^{2})$, where 0.45 is the asymptotic fraction of injected cosmic-ray energy transferred to neutrinos above $\gtrsim 10^{21}$~eV (as shown in Fig.~1 of Ref.~\cite{ESS}) and $E_t\sim 2 \times 10^{20}$~eV governs the onset of photopion production above $\sim 3 \times 10^{19}$~eV.  The total neutrino flux at Earth can be cast as
\begin{equation} 
\frac{dN_{\nu}}{dE_\nu} = 
\frac{c}{4\pi}
\int_0^{z_{max}} 20 \, \varphi(E^\prime_p) {\cal S}\!\left[ (1+z) \times E_p^\prime \right]
\frac{dE^\prime_p}{dE_\nu}
\frac{\mathcal{W}(z)}{dz/dt} dz\,,
\label{eq:nuflux}
\end{equation}
where $E_p^\prime= 20\,(1+z)\,E_\nu$ and the factor $20$ reflects the approximation of each daughter neutrino receiving about $1/20$ of the injected proton energy.  A more detailed derivation of this formula is given in Appendix~\ref{appB}.  The additional factor of $(1+z)$ in ${\cal S}$ accounts for the lowering of the photopion energy threshold as the CMB temperature increases at higher redshift.  This simple formulation is similar to the notation based on neutrino yield functions often used in prior studies~\cite{ESS, Ave:2004uj, PAST-NU} and provides a reasonably accurate neutrino spectrum which agrees rather well with the literature (in the energy range most interesting to UHE neutrino detectors), with a deviation from the simulated spectra not larger than the variations typically seen between such simulations. The overall normalization (set by $\varphi$) is again chosen such that the corresponding predicted cosmic-ray flux agrees with the measured cosmic-ray data at $10^{19}$~eV (where it is well-determined), as discussed in Section~\ref{UHECR}.

Given an UHECR source evolution, $\mathcal{W}(z)$, we can calculate the expected flux of cosmogenic neutrinos produced through photopion production on the CMB.  
While our simple, analytic method allows for a more transparent look at the effects of source evolution, it does not fully encompass the particle physics involved, particularly the low and high energy tails of the distributions of particle decays (which affect the low- and high-energy ends of the neutrino spectrum).  We utilize the publicly-available simulation package CRPropa~\cite{Armengaud:2006fx}, which uses the SOPHIA~\cite{Mucke:1999yb} code to handle particle processes, for this purpose.  We have made use of the analytical estimate described above and an extensive comparison to previous results presented in the literature (with similar parameters), to verify the results.
At the redshifts of greatest interest, $z\sim 0-4$, only cosmic rays with $E \gtrsim 5 \times 10^{19} / (1+z)$~eV can ever have the ability to produce neutrinos through the GZK process (even with a decreased photopion threshold).  At lower energies, photopion production can be facilitated by the cosmic infrared background (IRB), resulting in additional neutrinos of correspondingly lower energy~\cite{GZK-IR}.  Additionally, extragalactic magnetic fields may increase the path length of cosmic-ray propagation~\cite{Stanev:2000fb}.  However, due to the various uncertainties involved, we have chosen not include these effects.
%
%%%%%%%%%%%%%%%%%%%%%%%%%%%
%%%%%%%%%%%%%%%%%%%%%%%%%%%

Fig.~\ref{fig:gzk1} compares the resulting neutrino flux from the GRB model to those expected from QSO-like evolution and the SFH, assuming an $E^{-2}$ injection spectrum (with a sharp cutoff at $10^{21}$~eV).  The flux shown is the sum of all neutrino flavors (both electron and muon types are produced at the source), as a detector such as ANITA has a nearly equal taste for each flavor.  As can be clearly seen, the stronger evolution of the GRB model produces a larger neutrino flux than other models.  Also shown are the expected sensitivities of ANITA~\cite{Barwick:2005hn} (which began taking data in late 2006) and the proposed ARIANNA/SalSA detectors~\cite{ASK}. The flux predicted by strong GRB evolution, with parameters given by our metallicity-biased model, extends well into the reach of ANITA.  This presents an opportunity (within the next few years) to either confirm or significantly constrain the parameters of this model in a manner that would not be possible by cosmic-ray observations alone.

We can also consider the effect of varying the assumed parameters in the injection spectrum.  Since only cosmic rays with energies greater than $10^{19}$~eV will ever contribute significantly (even at high redshifts) to the neutrino flux, harder cosmic-ray spectra will result in larger fluxes.  Shown in Fig.~\ref{fig:gzk2} are the neutrino fluxes arising from cosmic-ray injection spectra of $\gamma = 1.5$, $2.0$, and $2.5$, assuming the strong
evolution. In particular, a $\gamma = 2.2$ spectrum might arise from relativistic shocks (see Ref.~\cite{Dermer-GZK} for details).  Since $\gamma \lesssim 2$ would be quite difficult to reconcile with cosmic-ray observations, this range can be regarded as an upper bound, which will soon be constrained by ANITA. Assuming a lower cutoff energy would be similar in effect to a softening of the spectral index.  Another interesting scenario is producing a softer spectral index ($\gamma > 2$), even if each GRB possesses an intrinsic $E^{-2}$ spectrum.  As proposed in Ref.~\cite{Kachelriess:2005xh}, if the distribution of the high-energy cutoffs of cosmic-ray sources follows a power law, then the overall spectrum that is observed will follow a power law with a different slope.  Such an argument may be particularly applicable to GRBs, as cosmic rays might be produced by a spectrum of sources ranging from LLGRBs to strong GRBs with very metal-poor progenitors.

%--------------------------------------------------------------------%
\section{Discussion and Conclusions}
The increase in our understanding of gamma-ray bursts can be traced to the improved capabilities now available to study these phenomena.  The ability to quickly and accurately localize a GRB has led to the establishment of a GRB-supernova connection and allowed for the study of the host galaxies in which these events occur.  Observations indicate that these hosts tend to be underluminous, star-forming and metal-poor.  A connection between a GRB and its host galaxy metallicity is not surprising in the context of the collapsar model, which requires rapidly-rotating stars that lack a H/He envelope (in order to be in accord with supernova observations).  These requirements can be satisfied by a metal-poor progenitor star.  Introducing a metallicity bias leads to an accelerated evolution of the cosmic GRB rate density (relative to the SFH), 
which allows for a better fit to recent Swift data. In models that attribute ultrahigh energy cosmic rays to gamma-ray bursts, this evolution provides the history of UHECR production in the universe.  If GRBs are to account for the observed cosmic-ray spectrum, they \textit{must} generate a flux of cosmogenic neutrinos.

A broad approach which utilizes all available means of observation will continue to be of great utility in unveiling the mysteries of gamma-ray bursts, particularly a relation to UHECR.  Further observations of GRB host galaxies, along with those of core-collapse supernovae (particularly Type Ib/Ic), will provide invaluable information concerning GRB progenitors.  A systematic study of host metallicities would be important for firmly establishing the GRB-metallicity anti-correlation, which, along with more accurate measurements of cosmic metallicity evolution~\cite{Jimenez:2006ea}, would allow for improved GRB rate-related calculations.

Direct observations of neutrinos from GRBs (e.g., by IceCube~\cite{Ahrens:2003ix} or a km$^3$ Mediterranean detector~\cite{Cuoco:2006qd}) would confirm the acceleration of protons to high (though not necessarily ultrahigh) energies.  In order to produce a detectable signal, a particularly strong burst, with measured gamma-ray fluence $\gtrsim 3 \times 10^{-4}$ erg cm$^{-2}$ (e.g., the recent GRBs 060928~\cite{GRB060928} and 061007~\cite{GRB061007}, which were observed to have fluences very near this threshold) is required~\cite{Dermer:2006bb}.  While the needed neutrino telescopes (which may also reveal the sources of Galactic cosmic rays~\cite{Kistler:2006hp}) are still under construction, future observations by Swift~\cite{Burrows:2005gf} should reveal similar bursts, in addition to determining the redshift-dependent GRB rate with unprecedented precision.

Cosmogenic neutrinos present a unique tool to examine the GRB-as-proton accelerator conjecture.  While the cosmic-ray spectrum that will be measured by Auger will allow for further assessments of the viability of prospective source models, the combined measurements of neutrinos and UHECR would break degeneracies between the various models, as discussed by Seckel and Stanev~\cite{Seckel:2005cm}.  It is well-known that strong source evolution can lead to observable cosmogenic neutrino signals; however, a physically motivated model was lacking.  The enhanced cosmogenic neutrino fluxes expected to result from strong GRB evolution will allow for this model to be tested in a novel fashion. We have produced this strong evolution naturally through metallicity dependence, however, our general result is also applicable to \textit{any} GRB model that attempts to explain the observed Swift redshift distribution with additional evolution.

The sensitivity afforded by ANITA will allow for near-term examination which may either affirm or, if the expected flux is not found, place substantial constraints upon the model parameters.  The lower detection threshold achieved by an ARIANNA-like detector will allow for a realistic opportunity to discriminate between evolution models.  Measurements of fluxes consistent with that expected from GRB evolution would provide compelling evidence for this model.  As the relatively young field of particle astrophysics continues to progress, it is quite an exciting possibility that metals, once relegated to the realm of pure astronomy, may, in fact, hold the key to the production of the highest energy particles in the universe.

%---------------------------------------------------------------------%
\textbf{Acknowledgments.---}%
We would like to thank John Beacom for many illuminating discussions;
Shin'ichiro Ando, John Cairns, Charles Dermer, Kris Stanek, Floyd Stecker, Casey Watson, and Eli Waxman for helpful suggestions; and Baha Balantekin, Jim Beatty, and Terry Walker for useful comments on the manuscript.
This work is supported by the National Science Foundation under CAREER Grant No. PHY-0547102 to JB, the Department of Energy grant DE-FG02-91ER40690, and The Ohio State University.  

\appendix

%--------------------------------------------------------------------
\section{GRB Redshift Distribution}
\label{appA}
In calculating the redshift distribution of gamma-ray bursts expected to be observable by Swift, we follow the model of Le and Dermer~\cite{Le:2006pt}, using similar, but somewhat simplified, notation.  This requires the assumption of a characteristic GRB gamma-ray energy output ($\epsilon_{\gamma}$) and a duration in the GRB rest frame ($\delta t$), additionally assuming flat temporal and spectral burst profiles.  Each burst is assigned a jet opening angle, $\theta$, which is selected from a power law distribution of the form $g_0 (1-\cos \theta)^s$, between an assumed $\cos \theta_{\rm max}$ and $\cos \theta_{\rm min}$, normalized with $g_0 = (1+s)/[(1-\cos \theta_{\rm max} )^{1+s} -(1-\cos \theta_{\rm min} )^{1+s}]$.

The observed energy flux from a GRB at a luminosity distance, $d_\ell (z)$, then takes the form
\begin{equation}
f_{}=\frac{\epsilon_{\gamma}/ \delta t}{4\pi d^2_{\ell} (1-\cos \theta) \lambda_b}
\end{equation}
where $\lambda_b \simeq 5$ is a bolometric correction factor, which accounts for the fraction of the burst spectrum within the energy band of the GRB detector.  Solving this equation for the detector trigger sensitivity threshold,  $f_{\rm det}$, yields the beaming angle, $\theta_{\rm det}$, required for detection.  In effect, by assuming a constant $\epsilon_{\gamma}$ we will use the opening angle as a proxy for the isotropic equivalent energy ($E_{\rm iso}$) often used in GRB studies (e.g., Ref.~\cite{Guetta:2003zp}).

For a comoving GRB rate, $n_{c} (z)$, the detectable GRB rate is reduced to $n_{c}(z) (1-\cos \theta)$ by beaming through angle, $\theta$.  We then find the detectable GRB fraction as
$$
\Theta(z) = \int_{\cos {\rm Min}[\theta_{\rm max},\theta_{\rm det}]}^{\cos \theta_{\rm min}} 
 (1-\cos \theta)\, g_0\, (1-\cos \theta)^s \, d\cos \theta 
$$
\begin{equation}
= 
\frac{1+s}{2+s} 
\frac{(1-\cos {\rm Min}[\theta_{\rm max} , \theta_{\rm det}])^{2+s} - (1-\cos \theta_{\rm min} )^{2+s}}{(1-\cos \theta_{\rm max})^{1+s} - (1-\cos \theta_{\rm min})^{1+s}}\,,
\label{a1}
\end{equation}
We then compute the number of detected bursts per unit redshift as
\begin{equation}
\frac{dN}{dz \, d\Omega} = \frac{1}{4\pi}\Theta(z) \frac{n_{c} (z)}{1+z} \frac{d V_c}{dz} dz\,,
\label{a2}
\end{equation}
where $1/(1+z)$ takes into account cosmic time dilation, and the comoving volume element is given as 
\begin{equation}
\frac{dV_c}{dz} = \frac{4 \pi\, d_\ell^2\,c}{(1+z) dz/dt} \, .
\label{a3}
\end{equation}
%
% with (as in Section~\ref{UHECR})
% \begin{equation}
% \frac{dz}{dt} = H_0\, (1+z)\, \left[\Omega_M\, (1+z)^3 +\Omega_\Lambda\right]^{1/2}\,
% \label{a4}
% \end{equation}
% where $\Omega_{M} = 0.3$, $\Omega_{\Lambda} = 0.7$, and ${H}_{0} = 70$~(km/s)/Mpc.  

Substitution of Eqs.~(\ref{a1}) and (\ref{a3}) into Eq.~(\ref{a2}) agrees with the formulation presented in Ref.~\cite{Le:2006pt}.

%-------------------------------------------------------------------------
\section{Cosmogenic Neutrino Spectrum}
\label{appB}
We describe a simple, analytic method to compute the cosmogenic neutrino spectrum observed at Earth, based upon the assumption that each daughter neutrino resulting from a photopion interaction receives $\sim 1/20$ of the parent proton energy.  That is, a neutrino \textit{detected} with energy, $E_\nu$, was produced with energy, $E_\nu^\prime = (1+z) E_\nu$, at the source, from a  parent proton which had energy, $E_p^{\prime}= 20\, E_\nu^\prime = 20\, (1+z)\, E_\nu$.  We predict the observed ($z=0$) neutrino spectrum, $ {dN_{\nu}}/{dE_{\nu}}$,by integrating over the contributions to the flux from redshifts up to $z_{\rm max}$ as
\begin{equation}
\frac{dN_{\nu}}{dE_{\nu}} = \frac{c}{4 \pi }
\int_0^{z_{max}} \frac{dN_{\nu}}{dE_{\nu}^\prime}  \frac{dE_{\nu}^\prime}{dE_{\nu}}\, 
\frac{\mathcal{W}(z)}{dz/dt} \,dz
\label{b1}
\end{equation}
where $dE_\nu^\prime/dE_\nu=(1+z)$ accounts for the fact that neutrinos are observed in a narrower energy bin then they were originally produced (due to redshifting).  Evolution in the density of sources is accounted for through $\mathcal{W}(z)$. The neutrino spectrum at production, ${dN_{\nu}}/{dE_\nu^\prime}$, can  be directly related to the cosmic-ray injection spectrum $\varphi(E_p^\prime) $ (per unit comoving volume per unit energy, per unit time), through the conservation of the transferred energy, as
\begin{eqnarray}
E_\nu^\prime \frac{dN_{\nu}}{dE_\nu^\prime} &=& 
\left( E_p^{\prime}\, \varphi(E_p^{\prime})  {\cal S}\!\left[ (1+z) \times E_p^{\prime} \right]  \right)
\frac{dE_p^{\prime}}{dE_\nu^\prime}  \\
\frac{dN_{\nu}}{dE_\nu^\prime} \frac{dE_\nu^\prime}{dE_\nu}
&=&  
20 \,\varphi(E_p^\prime) \,  {\cal S}\! \left[(1+z) \times E_p^{\prime} \right]  \, 
\frac{d E_p^{\prime} }{d E_\nu} 
\label{b2}
\end{eqnarray}
with ${\cal S}$ as defined in Section~\ref{NU}.  Substitution of Eq.~(\ref{b2}) into Eq.~(\ref{b1})
yields Eq.~(\ref{eq:nuflux}).

%---------------------------------------------------------------------%
%\textbf{References}

\end{document}